\documentclass[aps,a4paper,floatfix]{revtex4}

\usepackage[utf8]{inputenc}

\usepackage{graphicx,color}
\usepackage{amsmath,amssymb,amsfonts,amsthm,amscd,bm}
\usepackage{booktabs}
\usepackage{cancel}

\def\tilde{\widetilde}
\def\bar{\overline}
\def\hat{\widehat}
\def\*{\star}
\def\[{\left[}
\def\]{\right]}
\def\({\left(}      

\def\){\right)}

\def\zbar{{\bar{z} }}
\def\frac#1#2{\dfrac{#1}{#2}}
\def\inv#1{\dfrac{1}{#1}}

\def\d{\partial}

\def\2pi{\hbox{$2\pi i$}}

\def\dsl{\raise.15ex\hbox{/}\kern-.57em\partial}
\def\Dsl{\,\raise.15ex\hbox{/}\mkern-.13.5mu D}
\def\vep{\varepsilon}

      \def\CF{{\cal F}}

\def\CM{{\cal M}}   \def\CN{{\cal N}}   
      
\def\CS{{\cal S}}

\def\2pi{\hbox{$2\pi i$}}

\def\dsl{\raise.15ex\hbox{/}\kern-.57em\partial}
\def\Dsl{\,\raise.15ex\hbox{/}\mkern-.13.5mu D}
\font\numbers=cmss12
\font\upright=cmu10 scaled\magstep1
\def\stroke{\vrule height8pt width0.4pt depth-0.1pt}
\def\topfleck{\vrule height8pt width0.5pt depth-5.9pt}
\def\botfleck{\vrule height2pt width0.5pt depth0.1pt}
\def\Zmath{\vcenter{\hbox{\numbers\rlap{\rlap{Z}\kern
    0.8pt\topfleck}\kern 2.2pt
    \rlap Z\kern 6pt\botfleck\kern 1pt}}}
\def\Qmath{
    \vcenter{\hbox{\upright\rlap{\rlap{Q}\kern3.8pt\stroke}\phantom{Q}}}}
\def\Nmath{\vcenter{\hbox{\upright\rlap{I}\kern 1.7pt N}}}
\def\Cmath{\vcenter{\hbox{\upright\rlap{\rlap{C}\kern
                   3.8pt\stroke}\phantom{C}}}}
\def\Rmath{\vcenter{\hbox{\upright\rlap{I}\kern 1.7pt R}}}
\def\Z{\ifmmode\Zmath\else$\Zmath$\fi}
\def\Q{\ifmmode\Qmath\else$\Qmath$\fi}
\def\N{\ifmmode\Nmath\else$\Nmath$\fi}
\def\C{\ifmmode\Cmath\else$\Cmath$\fi}
\def\R{\ifmmode\Rmath\else$\Rmath$\fi}
\def\barray{\begin{eqnarray}}
\def\earray{\end{eqnarray}}
\def\beq{\begin{equation}}
\def\eeq{\end{equation}}

\def\Li{{\rm Li}}

\def\AA{\leavevmode\setbox0=\hbox{h}
\dimen0=\ht0 \advance\dimen0 by-1ex\rlap{\raise.67\dimen0\hbox{\char'27}}A}

\def\Li{{\rm Li}}

\makeatletter
\def\iddots{\mathinner{\mkern1mu\raise\p@
\vbox{\kern7\p@\hbox{.}}\mkern2mu
\raise4\p@\hbox{.}\mkern2mu\raise7\p@\hbox{.}\mkern1mu}}
\makeatother

\def\Li{{\rm Li}}

\def\Tbar{\bar{T}}

\def\vep{\varepsilon}

\def\cdd{{\rm cdd}}
\def\mr{\tfrac{mR}{2}}

\def\Li{{\rm Li}}
\def\Lr{{\rm Lr}}

\theoremstyle{plain}

\theoremstyle{remark}

\begin{document}

\title{
 $T \Tbar$ deformation of the Ising model and its ultraviolet completion}
\author{
 Andr\'e  LeClair\footnote{andre.leclair@gmail.com}
}
\affiliation{Cornell University, Physics Department, Ithaca, NY 14850} 

\begin{abstract}

Pure $T\Tbar$ deformations of conformal field theories are generally asymptotically  incomplete in the ultra-violet (UV) 
due to square-root singularities in the ground state energy on a cylinder of circumference $R$, such that the theory is ill-defined for distances shorter than  some critical 
$R_*$.    In this article we show how a  theory can be completed if one includes an infinite number of additional irrelevant perturbations.    
This is fully  demonstrated in the case of the Ising model at $c_{IR}= 1/2$ in the infra-red (IR),  where we find two completions with central charges
$c_{UV} = 3/2$ and $c_{UV} = 7/10$, the latter being the tri-critical Ising model.  Both of these UV completions  have $\CN=1$ supersymmetry which is broken in the
renormalization group flow to low energies.    We also consider multiple $T\Tbar$ deformations of a free massless boson,  where we cannot find a UV completion that is consistent with the c-theorem.     For negative coupling $g$,  which violates the c-theorem,  in both cases we find $c_{UV} = -c_{IR}$ as $g \to -\infty$.
Finally we also study pure $T\Tbar$ deformations of the off-critical Ising model.

\end{abstract}

\maketitle
\tableofcontents

\section{Introduction}

Our current understand of Renormalization Group (RG) flows and effective field theory makes an important distinction between relevant verses irrelevant operators. 
In the ideal scenario,  at very high energies the theory is governed by an ultra-violet (UV) fixed point of the RG, which is a conformal field theory.   At lower energies the behavior is
determined by relevant perturbations of the UV fixed point.    Relevant operators, which are usually finite in number,  generally introduce a mass scale and the spectrum consists of massive particles.
Under RG the theory flows to an  infra-red  (IR) fixed point described by a different conformal field theory (CFT).      If all the particles are massive, at low energies the particles effectively have infinite mass and decouple leaving an empty theory.   However,  if there are massless particles they can survive the flow to the deep IR such that the theory flows to a non-trivial IR fixed point.    In the vicinity of the IR fixed point the behavior is described by irrelevant perturbations of the IR 
CFT.   Perturbation of the UV fixed point has a high level of predictability since there are usually only a finite number of relevant operators and one can in principle predict the low energy properties.    However the reverse is not true:  there are an infinite number of irrelevant operators which introduces an infinite number of couplings in the low energy theory.
     In general it is impossible to reconstruct the UV behavior from what is known at low energies,  since some UV features are lost in the flow to the IR;  for example the massive fields that decoupled.    
One says that the RG flow to the IR is irreversible,  and this is captured for instance by the c-theorem \cite{ctheorem}.    This irreversibility is central to many of the   
difficulties  currently faced     in fundamental High Energy Physics and Quantum Gravity.   The so-called hierarchy problem in the Standard Model is of this nature. 
Also,  quantization of Einstein gravity involves a perturbation of free massless gravitons by  an infinite number of irrelevant operators.   In light of these remarks,  an interesting and fundamental question arises:  Is it possible to fully reconstruct the UV behavior from 
the properties of the IR fixed point CFT and its irrelevant perturbations?   In other words:   Is it possible to reverse the RG flow {\it without } violating the c-theorem and the like?  
Since based on our current understanding the answer is  likely to be No,  it would be very interesting to find concrete examples where the answer is Yes.   For the latter,  the theory is 
referred to as being  asymptotically UV complete.   In the context of quantum gravity this property of the RG flow is referred to as 
asymptotic safety \cite{Weinberg}.    In this article we will refer to this phenomenon as asymptotic UV reversibility, and will distinguish between such flows that violate the c-theorem and those that do not.

\def\mass{{\rm mass}}

A concrete framework in which to study the above profound  issues is in the context of so-called $T\Tbar$ deformations, i.e. perturbations,  of integrable quantum field theory in 2 space-time dimensions. 
The original works are are due to Smirnov and Zamolodchikov,  and Cavagli\`a et. al. \cite{ZTT,SZ,Tateo1}.   The models can be formulated with the action
\beq
\label{Action}
\CS = \CS_{IR} + \delta \CS
\eeq
where $\CS_{IR}$ is formally the action of the theory in  the deep  IR, and $\delta \CS$ is a perturbation by irrelevant operators.  
The theory $\CS_{IR}$ can be  taken to be a massive integrable theory or a CFT,  however as we will explain the case where $\CS_{IR}$ is an IR fixed point CFT is the most 
interesting.      
Of special interest is  the ``$T\Tbar$" perturbation, since it is typically the lowest dimension irrelevant operator.   
Every theory has a conserved stress-energy tensor, 
\beq
\label{Conv1}
T =-2 \pi \, T_{zz} , ~~~\Tbar = -2 \pi \, T_{\zbar\zbar}, ~~~ \Theta = 2 \pi\,  T_{z\zbar} , 
\eeq
where $z=x+i y$, $\zbar = x-i y$ are euclidean light-cone coordinates.    If $\CS_{IR}$ is a CFT then the trace of the stress-energy tensor 
$\Theta =0$.
It was shown in \cite{ZTT} that the dimension 4 irrelevant operator 
\beq
\label{Conv2}
T \Tbar = 4 \pi^2 \( T_{zz} T_{\zbar\zbar} - (T_{z \zbar})^2 \) 
\eeq
is well defined.  It was shown in \cite{SZ}  that the theory defined by   
\beq
\label{Conv3}
\delta \CS =  \frac{\alpha}{\pi^2} \int d^2 x \, T\Tbar
\eeq
continues to be integrable.   Here $\alpha$ is a coupling with dimension $1/\mass^2$.    (The $1/\pi^2$ in the normalization is chosen for  the convenience of simplifying the  Burgers differential equation below.)   It was also shown that the $T\Tbar$ perturbation leads to  an additional CDD factor 
$S_\cdd (\theta) $ in the two particle S-matrix where $\theta$ is the usual rapidity parameterizing the energy of a particle $E = m \cosh \theta$.   There has been a large amount of 
interesting subsequent work which focuses primarily on pure $T\Tbar$ deformations.     A partial list includes 
 \cite{Tateo1,Tateo2,Rosenhaus}.  
 These  models  are also of interest in connection to 2-dimensional gravity and string theory,  for instance in JT gravity;        the $T\Tbar$ perturbation may be 
viewed as a gravitational dressing of $S_{IR}$ \cite{Dubovsky,Dubovsky2,Cardy,Tateo3,Verlinde,Hartman,Frolov, Oku}.
For a review see \cite{Jiang}.

An important probe is the ground state energy $E(R)$ on an infinite cylinder of circumference $R$, which was studied in \cite{ZTT,Tateo1,SZ}.    In thermodynamic language the free energy density is
$\CF (T) =  E(R) / R$,  where $R=1/T$ is the inverse temperature. 
It is standard to express these quantities in terms of a scaling function $c(mR)$ 
\beq
\label{cR}
E(R) = - \frac{\pi}{6} \, c(mR)\,  /R
\eeq
where $m$ can be identified with a physical energy scale, such as the mass of a particle, or the energy scale of massless particles.  
The UV limit is $r \equiv  mR \to 0$,  whereas the IR corresponds to $r\to \infty$.    The normalization in \eqref{cR} is such that $c=1$ for a free massless boson.  
For a conformal theory,  $c(mR)$ is scale invariant, i.e. independent of $mR$,  and  for unitary theories is equal to the Virasoro central charge.  
(For non-unitary theories it is shifted $c\to c -12 d_0$ where $d_0$ is the lowest scaling dimension of fields.)
The quantity $c(mR)$ can be used to track the RG flow.
In our recent work we studied the quantity $c(mR)$ with methods different from those in \cite{SZ,Tateo1},  namely with the Thermodynamic Bethe Ansatz (TBA)  \cite{AL}.  
The TBA was also studied recently in this context for more complicated supersymmetric models in \cite{Ebert}.  
Specifically we took $\CS_{IR}$ to be a free massless boson or fermion and varied the $IR$ central charge $c_{IR}$ by varying the chemical potential.  
The S-matrix is then a pure CDD factor
\beq
\label{Scdd} 
S_\cdd (\theta ) = e^{i g \sinh \theta }
\eeq
where $g$ is a dimensionless coupling.    From the TBA we found the general result 
\beq
\label{chGeneral}
c(h) = \frac{2 c_{IR}}{1 + \sqrt{1- \frac{2 \pi h}{3} c_{IR} }} , ~~~~~h \equiv \frac{g}{(mR)^2}
\eeq
where
 $c_{IR}$ is a constant $-\infty < c_{IR} < \infty$ identified as the IR central charge as $R \to \infty$, and $m$ is the energy scale in the massless TBA.   We can compare with the 
 analogous result obtained in \cite{SZ,Tateo1} based on the inviscid Burgers equation
 \beq
 \label{Burgers}
 \d_\alpha E + E \d_R E = 0 .
 \eeq
 Indeed, one finds that \eqref{chGeneral} satisfies the above differential equation if one identifies
 \beq
 \label{params}
 h = - \frac{\alpha}{R^2}, ~~~~~\Longrightarrow ~~ g = - \alpha \, m^2.
 \eeq
 This successful comparison is an indication that the massless TBA equations proposed in \cite{AL}, which involved factorizing the CDD factor, are indeed correct.   We will review this proposal in Section V.

 The issue of UV completeness is distinct from integrability.  
 Based on the expression \eqref{chGeneral},  in \cite{AL} we reached the following conclusions concerning the issue of UV completeness.    
 First of all,  the c-theorem states that $c(mR)$ increases toward the UV, and this requires $g>0$, i.e. $\alpha < 0$.    For $c_{IR} < 0$ the flow is well-defined for arbitrarily small $R$, i.e. in the limit 
 $h \to \infty$.   In this case the theory was interpreted as being UV complete with central charge $c_{UV} =0$.   However the fact that $c_{UV}=0$ for {\it all}  initial 
 $c_{IR}<0$ seems unsatisfactory;   this issue will be resolved below.   
 On the other hand,  for the physically more interesting case of positive $c_{IR}$,  $c(mR)$ develops a square-root singularity at 
 $h= h_* \equiv  3/(2 \pi c_{IR})$.    At the singularity $c(h_*) \equiv c_{UV_*}= 2 c_{IR}$ is finite.     The flow cannot be extended to arbitrarily small $R$ since $c(mR)$ becomes complex for $h> h_*$,  and one should conclude that the
 theory is {\it not} UV complete.    The above discussion is for $g>0$ since that is what is consistent with the c-theorem,   however $g<0$ is still interesting as discussed in 
 \cite{AL},  and we will return to this below.   
 
 Having made these introductory remarks,  we can now state the primary findings of our work presented below.   
 By including an infinite number of additional irrelevant perturbations, and tuning the couplings appropriately, 
 the theory can be asymptotically completed with a predictable UV fixed point with central charge $c_{UV}$.   
 We demonstrate this in the following specific context.   For concreteness, first  consider $\CS_{IR}$ as defining a massive integrable QFT.   
 Integrability implies an infinite number of local conserved currents of spin $\pm (s+1)$ where $s$ is a positive integer.   From these currents one can construct 
 left/right bilinear local fields  $X_s$  of dimension $(\mass)^{2(s+1)}$, where $X_1 = T\Tbar/\pi^2$.     It was shown by Smirnov and Zamolodchikov that perturbation by all of these operators 
 \beq
 \label{delS}
 \delta \CS = \sum_{s \geq 1} \alpha_s \, \int d^2 x \, X_s (x)
 \eeq 
 continues to be formally integrable for any choice of couplings $\alpha_s$  (including $\alpha_s =0$ for some $s$).  
 Below we show that by choosing the couplings $\alpha_s$ appropriately, the theory is asymptotically complete in the UV, i.e. the RG flow extends to arbitrarily short distances 
  where it becomes a different CFT with central charge $c_{UV}$.   If the flow is consistent with the c-theorem,  then  $c_{UV} > c_{IR}$.   Our analysis is based on the TBA which is 
  particularly well suited to the study of these issues since  we know the general form of the CDD factor.    
  We do not demonstrate this in full generality, but rather in an example that is non-trivial enough for our purposes:  $T\Tbar$ perturbations of the critical or non-critical Ising model
  where $c_{IR}=1/2$.    We show how to choose the couplings $\alpha_s$ such that $c_{UV} = 3/2 $ or $7/10$, the latter  corresponding to the tri-critical Ising model.  
  The fact that the UV completions are not unique is to be expected since there is some freedom in choosing the couplings $\alpha_s$.

In the next section we first review the construction of the operators $X_s$ \cite{SZ}.     In Section III we turn to the example of the Ising model, focusing on the massless critical case.
The massive non-critical case with only $\alpha_1 \neq 0$ can also be dealt with in rather great detail as described in section VI.

\section{Perturbations  by $T\Tbar$ and other irrelevant operators based on higher integrals of motion}

In this section we review the main results presented in the pioneering work \cite{ZTT,SZ,Tateo1}.   
Let us first consider the case where $\CS_{IR}$ corresponds to a massive integrable model.   
Then there exists an infinite number of conserved local currents satisfying the continuity equations
\beq
\d_\zbar T_{s+1} = \d_z \Theta_{s-1}, ~~~~~~
\d_z \Tbar_{s+1} = \d_\zbar  \bar{\Theta}_{s-1},
\eeq
where $s$ is a positive integer 
with $s+1$ and $-(s+1)$  the spins of $T_{s+1}$ and $\Tbar_{s+1}$ respectively.  
For $s=1$ these are the components of the stress-energy tensor and the conserved charges are left and right components of momentum.   
The spectrum of the integers $\{ s \}$ depends on the model in question.   For models based on 
${\rm su} (2)$ such as the sine-Gordon or sinh-Gordon models, $s$ is an odd integer.  
Zamolodchikov showed that from these one can construct well defined local operators $X_s$:
\beq
\label{Xsdef}
X_s = T_{s+1} \Tbar_{s+1} - \Theta_{s-1} \bar{\Theta}_{s-1}
\eeq
with scaling dimension $(\mass)^{2(s+1)}$.    More importantly, perturbation by such operators preserves the integrability.    
Thus we consider the theory defined by the action 
\beq
\label{CSgeneral}
\CS = \CS_{IR} + \sum_{s\geq 1}  \alpha_s \, \int d^2 x \, X_s (x) 
\eeq
where $\alpha_s$ are coupling constants of scaling dimension $(\mass)^{-2s}$.   Our convention for $\alpha_1$ is that
$X_1 = T \Tbar/\pi^2$.    We will refer to the above multi-parameter perturbations simply as $T\Tbar$.

Since the theory is integrable, it can be described by a factorizable scattering theory for some fundamental particles.  For simplicity we suppose the spectrum consists of a single particle  of mass $m$ with energy and momentum parameterized as usual by the rapidity $\theta$:
\beq
\label{rapid}
E = m \cosh \theta, ~~~~~ p = m \sinh \theta, 
\eeq
and the S-matrix for the unperturbed theory is $S_{IR}( \theta)$.   Then the effect of the perturbation is to multiply the S-matrix by a CDD factor
\beq
\label{SScdd}
S (\theta) = S_{IR} (\theta) \cdot S_\cdd (\theta),  ~~~~~~~
S_\cdd (\theta) = \exp \( i  \sum_{s \geq 0} g_s \,\sinh (s \theta) \).
\eeq
The normalization of the operators $X_s$ can be chosen such that 
\beq
\label{gs}
g_s = - \alpha_s \, m^{2s} ,
\eeq
and the convention for $\alpha_1 = \alpha$ is the same as in \cite{SZ}.   

We already encounter a serious difficulty:   The infinite series in $S_\cdd$ has very little domain of convergence.   As we will explain this can be resolved if one first takes a massless limit and then factorizes $S_\cdd$,  as was proposed for the pure case in \cite{AL}.

\section{Asymptotic reversibility of the $T\Tbar$ perturbed Ising model}

\subsection{Generalities}

Whether the unperturbed theory is massive or massless is clearly an IR property.   At very high energies compared to the physical mass of particles, the theory is effectively massless.  
For this reason,  as far as the issues raised in the Introduction that we wish to address,  there is little to be gained by studying  $T\Tbar$ perturbations of massive theories.  
For instance,  a pure $T\Tbar$ perturbation where only $g_1 \neq 0$ of a massive theory is expected to have the same UV singularity as the perturbation of the massless case since the behavior in the UV is dominated by irrelevant operators.   We will show this explicitly  for an example in the next section.   
A precise statement is that a massive spectrum is associated with a relevant operator whereas in the ultra-violet the behavior is controlled by the irrelevant operators since the latter are less and less important at low energies,  hence the terminology ``irrelevant".   If both relevant and irrelevant operators are present,  they compete:   the relevant operators win at low energies whereas the irrelevant ones win at high energies.     However a massive theory plays an indirect role if one is interested in CFT's perturbed by  irrelevant operators since a relevant perturbation selects a spectrum of particles that describes the IR CFT which in turn affects the analysis of the UV properties.  
Namely,  a massive theory selects a spectrum of particles and a scattering description of the IR CFT in terms of Left-Left and Right-Right S-matrices
$S_{LL}$ and $S_{RR}$,  and this in turn determines the operators $X_s$.     The formulation of massless factorized scattering in \cite{ZZ,Fendley} thus plays a central role.    It is important to recognize that the massless scattering description of a CFT is not unique and this implies that the UV properties of $T\Tbar$ deformations  are also not unique,  since the operators $X_s$ themselves depend on  the IR description of the IR CFT.  On the other hand every CFT has a stress-energy tensor and a resulting $X_1$,  thus there is simply not enough information in  a  pure $T\Tbar$ deformation to determine a UV completion.   
For instance the critical Ising model has two integrable perturbations, either by the energy operator,  which is just a mass term for the Majorana fermion,  or by the spin field.   
The latter requires working  with $8$ fundamental massless  particles related to the Lie algebra $E_8$ as discovered by A. Zamolodchikov.    For a comprehensive  review
of integrable perturbations of CFT,  the TBA, etc.,  we refer to the book by Mussardo \cite{Mussardo}.   

\bigskip

In summary,  we define a $T\Tbar$ perturbation of a CFT$_{IR}$ by the following steps:

\bigskip
\noindent (i) ~~ Begin with a CFT$_{IR}$ which will eventually be identified as the IR fixed point of the $T\Tbar$ flow.    Before turning on $T\Tbar$, we first consider a relevant perturbation of CFT$_{IR}$  which defines an integrable massive theory with a known  spectrum of particles,  their S-matrices,  and the operators $X_s$.  

\medskip
 \noindent (ii) ~~ Second, we  take the massless UV limit of the theory in (i)  to define a scattering description of the original CFT$_{IR}$ in terms of  scattering matrices $S_{LL}$ and $S_{RR}$.   The value of $c_{IR}$ is built in from the beginning and depends on the spectrum and S-matrices $S_{LL}, S_{RR}$.    
 
 \medskip
 \noindent (iii) ~~  Third,  we turn on $X_s$ perturbations of the CFT$_{IR}$, where the latter  is now the  IR fixed point of the $T\Tbar$ flow.    Note that CFT$_{IR}$ was the UV limit in step (ii), but now  serves as the IR fixed point of the $T\Tbar$ deformation.    For this reason the order of UV limits matters.  
  
 \medskip
 \noindent (iv) ~~ Finally we study whether the $T\Tbar$ perturbation is completed in the UV by a different CFT$_{UV}$.     
 
 \bigskip
 
 We itemized these steps in detail since the two UV limits involved in the procedure do not obviously commute.  In this section we assume we first take the UV limit which leads to a massless scattering description of CFT$_{IR}$ before turning on $T\Tbar$.

The UV issues we wish to resolve are already present for non-interacting theories $\CS_{IR}$ where in the massless limit $S_{LL} = S_{RR} = \sigma = \pm 1$, where 
$\sigma = 1,-1$ corresponds to bosonic verses fermionic quantum statistics.    For massive interacting theories the statistics is always fermionic since the bosonic case is unstable
 \cite{Simon}.   However in the present context bosonic statistics cannot be disregarded since the $T \Tbar$ interactions are too soft to modify the statistics.   
 For instance, both the bosonic and fermionic cases were needed in \cite{AL} in order to cover the full range of $c_{IR}$.       For this reason we will display $\sigma$ in many TBA formulas even though our primary example is
 the fermionic Ising model, which we now turn to in detail.

\def\psibar{\bar{\psi}}

\subsection{The Ising case}

The Ising model at its critical temperature $T_c$ is known to be described by a massless Majorana fermion with fields $\psi, \psibar$, 
which is a conformal field theory with $c=1/2$.    Perturbing the temperature away from $T_c$ corresponds to a mass term with 
mass $m \propto (T-T_c)$.   The integrals of motion are known to exist for $s$ an odd integer.   We thus consider a $T\Tbar$ perturbation defined by the action
\beq
\label{IsingT}
\CS = \inv{4 \pi} \int d^2 x \( \psi \d_\zbar \psi + \psibar \d_z \psibar +  m \, \psibar \psi  \) + \sum_{s \geq 1, ~{\rm odd}}  \alpha_s \int   d^2 x  \, X_s .
\eeq
As discussed above, 
we first take the UV limit $m\to 0$,  then consider the $T\Tbar$ perturbations of the $c=1/2$ free massless Majorana fermion.  

The massless TBA requires distinguishing between Left (L) and Right (R) movers,  where the energy and momentum of a  single particle is 
\barray
{\rm right ~ movers:} ~~~~~ E &=&  p= \tfrac{m}{2} e^{\theta} \cr
{\rm left  ~ movers:} ~~~~~ E &=& - p= \tfrac{m}{2} e^{-\theta}  .
\earray
The parameter $m$ is  needed to give $E,p$ units of energy. 
Here $m$ is not the mass of a physical particle,  but is a physical mass scale arising from the dimension-full parameters $\alpha_s$ as in \eqref{params}.    
Since the unperturbed theory has S-matrices $S_{LL} = S_{RR} = \sigma = -1$,  the TBA equations just couple the L,R pseudo-energies $\vep_{L,R}$ and have the standard form 
\barray
\vep_R (\theta) &=&  \mr  e^{\theta} + \sigma  \,G_{RL} \star \log \( 1-\sigma \, e^{-\vep_L (\theta)} \)  \cr
\vep_L (\theta) &=&  \mr  e^{- \theta} + \sigma \,  G_{LR} \star \log \( 1-\sigma \, e^{-\vep_R (\theta)} \) ,  
\label{TBAmassless}
\earray
 where we have defined the convolution 
\beq
\label{convo}
\( G \star f \) (\theta) = \int_{-\infty}^\infty \frac{d\theta'}{2\pi} \, G(\theta - \theta') f(\theta') 
\eeq
for an arbitrary function $f(\theta)$.    
Finally,
\beq
c(mR) = c_L (mR) + c_R (mR)
\eeq
where 
\barray
c_R &=&  - \frac{3\sigma}{ \pi^2} \,  \int_{-\infty}^\infty d\theta  \,  \tfrac{mR}{2} \, e^{\theta}  \, \log\( 1-\sigma\, e^{-\vep_R (\theta) } \)  \cr 
c_L &=& -  \frac{3\sigma}{ \pi^2} \,  \int_{-\infty}^\infty d\theta   \, \tfrac{mR}{2} \,  e^{- \theta}  \, \log\( 1-\sigma\, e^{-\vep_L (\theta) } \)  .
\label{cLR}
\earray

It remains to determine the kernels $G$,  which must follow from the CDD factor $S_\cdd$.   For the pure case, we previously proposed that it is necessary
to factorize the CDD factor \cite{AL}.    The essential reason is that if one chooses 
$G_{RL} = G_{LR}= G(\theta) = -i \d_\theta \log S_\cdd (\theta)$, then the convolution integrals $\star$ do not converge in  an iterative solution, and the same is true for 
multiple $T\Tbar$ perturbations.   A justification for this factorization is that it leads to results that agree with the solutions that follow from the Burgers equation, as discussed in the Introduction.    The necessary factorization is 
\beq
\label{LR}
S_\cdd (\theta) = S_{LR} (\theta) S_{RL} (\theta),  ~~~~~~~~~~ 
S_{RL} (\theta ) =  \exp \( i \sum_{s\geq 1}  g_s  e^{s\theta} /2 \), ~~S_{LR}  (\theta) = \exp \( -i \sum_{s \geq 1}  g_s e^{-s\theta} /2 \) . 
\eeq
The kernels which follow from these S-matrices, $G_{RL} (\theta) = -i \d_\theta \log S_{RL} (\theta)$ and similarly for $G_{LR}$, are 
\beq
\label{GLR} 
G_{RL} (\theta) = G_{LR} (-\theta) =  \sum_{s \geq 1}  s \, g_s  \, e^{s\theta} /2 .
\eeq
The $\theta \to -\theta$ symmetry of the TBA equations implies
\beq
\label{LRsym}
\vep_L (\theta) = \vep_R (-\theta), ~~~~~ \Longrightarrow ~~ c_L = c_R.
\eeq

\def\Ghat{\hat{G}}

The above factorization can resolve the problem of the convergence of the multiple CDD factor itself.    The kernel
\beq
\label{Gmassive}
G(\theta) = -i \d_\theta \log S_\cdd (\theta)  = \sum_{s\geq 1} s g_s \, \cosh (s \theta), 
\eeq
has a very limited domain of convergence.   On the other hand, suppose that the couplings $g_s$ are such that the kernels in \eqref{GLR} converge  to a well-defined function 
$\Ghat^-$ for 
$\theta < 0$:
\beq
\label{Ghat}
G_{RL} (\theta) = G_{LR} (-\theta) = \Ghat ^- (\theta) ~~~~~~ {\rm for} ~~ \theta < 0.
\eeq
If the function $\Ghat^-$ extends to positive $\theta$ via  $\Ghat^-(\theta) = \Ghat^-(-\theta)\equiv \Ghat (\theta)$, then 
for all $\theta$ we define the kernels from this function 
\beq
\label{Alltheta}
G_{RL} (\theta) = G_{LR} (\theta) = \Ghat (\theta) = \Ghat (-\theta). 
\eeq
The above equation ensures \eqref{LRsym}.  The precise meaning of this construction will be clear in the next subsection.

\subsection{Possible UV completions of Ising with $T\Tbar$} 

\def\Ghat{\hat{G}}
\def\Shat{\hat{S}}

Since our main purpose is to resolve the UV singularity found in the pure $T\Tbar$ perturbation,  we chose couplings 
such that they agree with previous conventions for $X_1$,  namely $g_1 = g = -\alpha \, m^2$ where $g$ is a continuous variable.
   (See the Introduction for definitions of $g$ and $\alpha$.)
We know that for the Ising model $X_s$ exist for $s = 2n+1$ a positive odd integer.    As we will explain, an  important criterion  for the existence of a UV fixed point 
is that the following integral converges
\beq
\label{Ghatconv}
\int_{-\infty}^\infty \frac{d\theta}{2 \pi} \, \Ghat (\theta) \equiv k.
\eeq
Note that for a pure $T\Tbar$ perturbation with only $g_1$ the above integral does not converge,  which essentially explains why the pure theory is not UV complete.

A very natural choice of couplings is the following:
\beq
\label{gss}
g_{2n+1} =  \frac{g\,(-)^n}{2n+1}, ~~~~~ n \geq 0.
\eeq
Then the kernels in \eqref{GLR} converge for $\theta < 0$ and $\Ghat^-$ can be identified with $1/\cosh \theta$.     
This leads to 
\beq
\label{Ghat}
\Ghat (\theta) = \frac{g}{4} \, \inv{\cosh \theta}.
\eeq
The integral in \eqref{Ghatconv} now converges and one finds
\beq
\label{kg}
k = g/8.
\eeq

The above kernel is associated with an  S-matrix $\Shat$:
\beq
\label{Shat}
\Ghat (\theta) = - i \d_\theta \, \log \Shat (\theta),    ~~~~~~
\Shat (\theta) = (S_0 (\theta) )^{g/4}, 
\eeq
where
\beq
\label{S0}
S_0 (\theta) = -i \tanh \( \tfrac{\theta}{2} - \tfrac{i\pi}{4} \).
\eeq
This suggests that $g$ equal to $4$ times an integer is special,  and this will indeed turn out to be the case.

The computation of the UV central charge $c_{UV}$ is standard and reviewed in the Appendix.   For convenience let us repeat the main formulas here.
We have 
\beq
\label{cless5f}
c_{UV} = \frac{6\sigma}{\pi^2} \( 2  \, \Lr_2 \( \sigma \, e^{-\vep_0} \) -  \Li_2 (\sigma) \), 
\eeq
where $\vep_0$ a solution to the transcendental equation 
\beq
\label{transf}
\vep_0 = \sigma k \log \( 1-\sigma e^{-\vep_0} \).
\eeq
Above, $\Li$ is the dilogarithm and $\Lr$ the Rogers dilogarithm,  
\beq
\label{rogersf}
\Lr_2 (z) =  \Li_2 (z) + \tfrac{1}{2}  \log |z| \log (1-z) .
\eeq

For $k>1$, i.e. $g>8$, the equation \eqref{transf} has no real solutions.  
As $g\to 8$, $\vep_0 \to - \infty$ and 
\beq
c_{UV} = 3/2 ~~~~~~~~~~{\rm when} ~~ g=8.
\eeq
On the other hand,  the equation has solutions for all $g<8$, including negative.   
One finds
\beq
\label{Lambert}
z_0 = e^{-\vep_0} \approx - \frac{W(-k)}{k}, ~~~~~~~~ k = g/8 \to -\infty
\eeq
where $W$ is the Lambert W-function.  
In this limit $z_0 \to 0$ and $\Lr (0) = 0$.   Thus
\beq
\label{cUVminus}
c_{UV} = - c_{IR} = -1/2 ~~~~~{\rm as} ~~ g \to - \infty
\eeq 
This is interesting since for the pure $T\Tbar$ case,  $c_{UV} = 0$ as $g \to -\infty$ for all $c_{IR}$ \cite{AL},  whereas the above result at least distinguishes between different
$c_{IR}$.      

In Figure 1 we plot $c_{UV}$ as a function of $k$.   One sees the feature anticipated in \cite{AL}:  for $g>0$, $c(r)$ increases toward the UV $r\to 0$, i.e. is consistent with the c-theorem,  however for $g<0$ the c-theorem is violated,  i.e. $c$ decreases toward the UV.     We thus conclude that based on our proposed kernel $\Ghat$, 
\beq
\label{cuvrange}
-1/2 \leq c_{UV} \leq 3/2 .
\eeq
Our analysis thus far does not at all ensure that there is a complete formulation of these flows as relevant perturbations of a known UV CFT with the above value of $c_{UV}$,  especially since  
the S-matrix $\Shat$ is peculiar for $g/4$ not equal to an integer.   

\begin{figure}[t]
 \label{cNeg}
\centering\includegraphics[width=.6\textwidth]{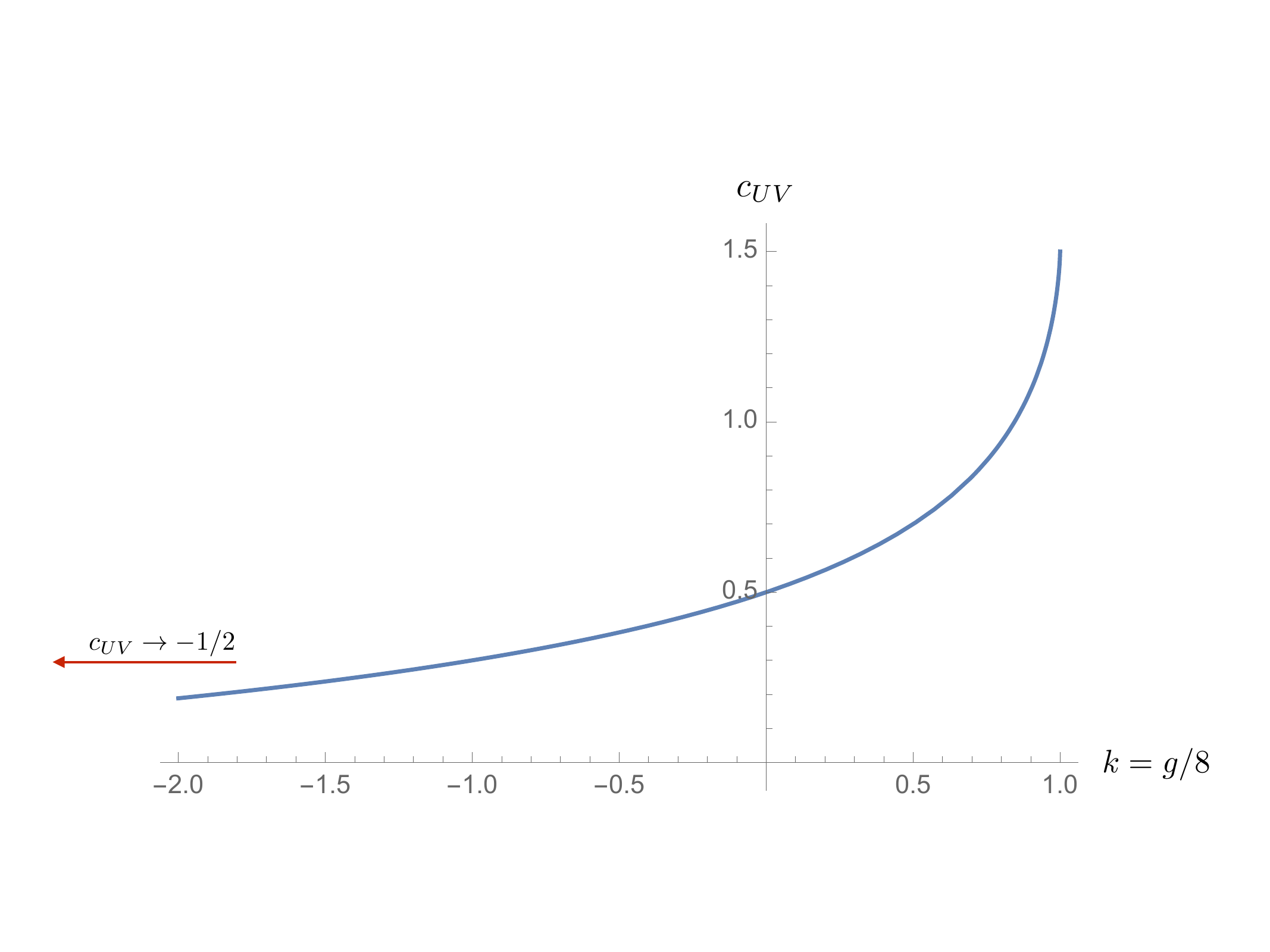}
\caption{$c_{UV}$ as a function of $k=g/8$ for $T\Tbar$ perturbations of the critical Ising model.  For $k>1$ there is no solution to the 
transcendental equation \eqref{transf}.}  
\end{figure}

There is one more interesting case which serves as a check of our construction.    
As noted above, $g=4$ is also special since $\Shat$ is a single power of $S_0$.   One finds from the above formulas 
that $z_0$ is the golden ratio:
\beq
\label{golden}
z_0 = \frac{1+\sqrt{5}}{2}
\eeq
and 
\beq
\label{710}
c_{UV} = \frac{7}{10} ~~~~~~~~~~~~~~  (g=4),
\eeq
which is the central charge of the  tri-critical Ising model,  the next model in the minimal unitary series. 

It turns out that the $g=4$ case was already known from the top down \cite{Alyosha}, however we ``discovered" it purely from the IR data.  
 In a larger context, 
 let $\CM_p$ denote the $p$-th  unitary minimal model of CFT with 
$c= 1- \tfrac{6}{(p+2)(p+3)}$ where $p=1,2,\ldots$.   The Ising model is $\CM_1$ and $\CM_2$ is the tri-critical Ising model.  Consider the integrable perturbation of $\CM_p$ by the relevant operator $\delta \CS = \lambda \int d^2 x \, \Phi_{1,3}$ where
``$\Phi_{1,3}$" has scaling dimension $\tfrac{2(p+1)}{(p+3)}$.   For $\lambda <0$ the theory is massive,  however it is massless for $\lambda >0$.  The latter flows to 
$\CM_{p-1}$ in the IR where it arrives there via the irrelevant operator $\Phi_{3,1}$  which has dimension $\tfrac{2(p+4)}{(p+2)}$ in $\CM_p$ \cite{ZMinimalFlow,CardyLudwig}.     
For the Ising model $\CM_1$,  the operator $\Phi_{3,1}$ does not exist,  and rather the flow from the tri-critical Ising model to the Ising model actually 
arrives via the operator $T\Tbar$ \cite{Alyosha}.

We have shown that  there are at least  two different UV completions of Ising for the class of kernel $\Ghat$ we have considered where  $g/4 = 1$ and $2$. 
It is interesting to note that these two most interesting cases  are both related to $\CN =1$ supersymmetry.   The theory with $c_{UV} = 3/2$ can be considered as a free boson plus a free Majorana fermion, which is a supersymmetric theory.    Apparently there exists a relevant perturbation of this theory that breaks the SUSY and in the flow to the IR
the boson becomes decoupled leaving only the Majorana fermion.   Also,  $c_{UV} = 7/10$ is the lowest member of the $\CN =1$ SUSY minimal CFT's.   
What is conceptually interesting about these flows is that, whereas the single fermion IR theory shows no signature of SUSY,   we were able to restore the broken SUSY in the flow to the UV,   whereas one is normally interested in a top down approach.   
This implies there is a hidden non-linear SUSY in the Ising model perturbed by $T\Tbar$.    Al. Zamolodchikov interpreted the
massless Majorana fermion as the goldstino of the spontaneously broken $\CN=1$ SUSY \cite{Alyosha}.    

Recall that the instability of the pure $T\Tbar$ perturbation occurs at $c_{UV_*} = 2 c_{IR} = 1$,  which is between $7/10$ and $3/2$.    Thus the UV completion with 
$c_{UV} = 7/10$ occurs before the instability is reached,  whereas the other completion occurs beyond at even shorter distances.

\section{Remarks on the bosonic case}

In this case the equation \eqref{transf} has no solutions for $g>0$,  thus there is no resolution of the square-root singularity in the UV,  and thus no evidence of a UV completion
based on our kernel $\Ghat$.   On the other hand  there are solutions for all $g<0$.    As $g\to -\infty$ the formula \eqref{Lambert} still applies for the bosonic case, 
and \eqref{cUVminus} still applies where now $c_{UV} = -1$.   Thus 
\beq
\label{BoseRange}
-1 \leq c_{UV} \leq 1
\eeq
and all these flows violate the c-theorem.   See Figure 2.

\begin{figure}[t]
 \label{cNeg}
\centering\includegraphics[width=.6\textwidth]{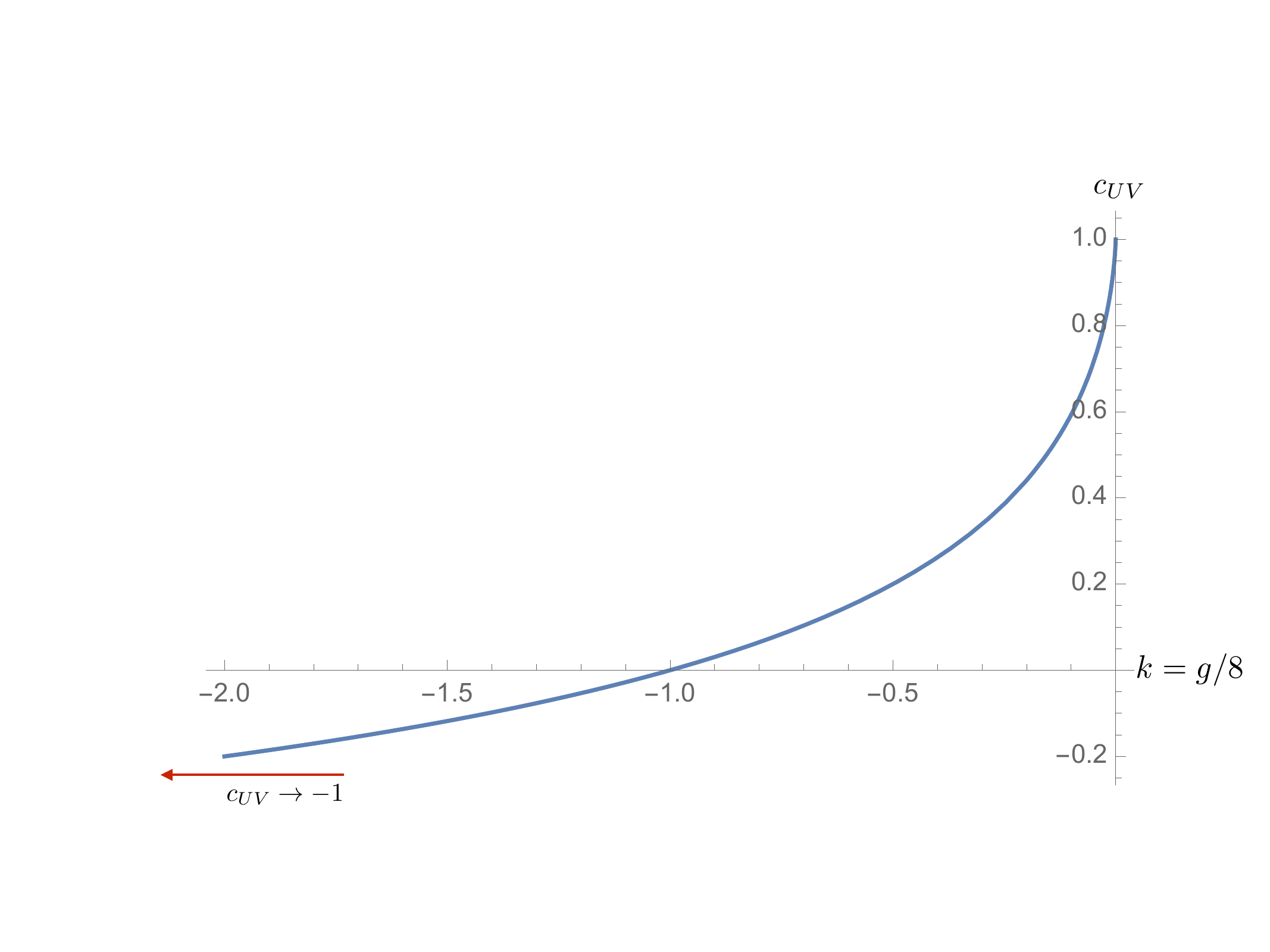}
\caption{$c_{UV}$ as a function of $k=g/8$ for $T\Tbar$ perturbations of the free massless boson.  For $k>0$ there is no solution to the 
transcendental equation \eqref{transf}.}  
\end{figure}

\section{Pure $T\Tbar$ for the off-critical Ising model and free massive boson.}

In this section we consider pure $T\Tbar$ deformations  of free {\it massive} theories where only $g_1 =g = -\alpha \, m^2$ is non-zero.  
As stated above,  there is not a great deal to learn conceptually from these cases,  however we present them here for a few reasons.  
First, this allows us to see explicitly that,  as expected,  these theories have the same UV singularity as the massless case since a non-zero mass is an IR property.  
Second,  these examples clarify the discussion in Section IIIA:   for multiple $X_s$ perturbations one must first take a massless limit in the IR otherwise the 
kernel in \eqref{Gmassive} in general does not converge.    Third,  it turns out that the resulting TBA equations can be solved exactly in a way similar to the massless case 
presented in \cite{AL}.

Suppose a theory consists of a single particle of mass $m$. 
The pseudo-energy $\vep (\theta )$  is now a  solution to the single integral equation:
\beq
\label{TBAmass}
\vep (\theta) = mR  \, \cosh \theta  +\sigma G \star \log\( 1-\sigma e^{-\vep} \).
\eeq
The kernel $G$ is 
$G(\theta) = -i \d_\theta \log S(\theta)$,
and  the scale dependent central charge is 
\beq
\label{cmass}
c(mR) = - \frac{3\sigma}{\pi^2} \int_{-\infty}^\infty d\theta ~ mR \cosh \theta \, \log \( 1-\sigma e^{- \vep (\theta)} \) .
\eeq

If the unperturbed theory is a free massive boson or fermion then the S-matrix is due entirely to the CDD factor,  with kernel 
\beq
\label{newG}
G(\theta)  = G_\cdd (\theta) = g \cosh \theta .
\eeq
Due to the above simple form of the kernel,  the solution to the TBA equation is greatly simplified. 
Let us first carry out the bosonic $\sigma = +1$ case. 
  One can express 
\beq
\label{epMass}
\vep (\theta) = \( 1 + h \, B\) r \cosh \theta
\eeq
where 
\beq
\label{rh}
r\equiv mR, ~~~~~ h \equiv \frac{g}{r^2}\,  .
\eeq
Plugging this expression into the TBA integral equation, $B$ is a function of $r,h$ and satisfies the integral equation
\beq
\label{Binteq}
B = \int_{-\infty}^ \infty \frac{d\theta}{2 \pi} \, r \cosh \theta \, 
\log\[ 1 - e^{-(1+h\, B)\, r \cosh \theta } \] .
\eeq
The scale dependent $c$ has the simple expression
\beq
\label{crh}
c(r,h) = - \frac{6}{\pi} \, B(r,h) .
\eeq
The equation \eqref{Binteq} is much easier to solve than a general TBA equation since $B$ is just a constant independent of $\theta$.   

\def\mless{\tilde{m}}

For non-zero mass
one must still solve \eqref{Binteq} numerically,   unlike the massless case  presented in \cite{AL}.   Let us show  how to obtain the latter result from the above formulas.  
The massless limit is taken as follows.  Make a change of variables $\theta \to \theta + a$ and perform the limits
\beq
\label{masslessLimit}
\lim_{m\to 0, a \to \pm \infty} = 
\begin{cases} 
\tfrac{\mless}{2} \, e^\theta , ~~~~~\mless = m e^a~ ~~{\rm held ~ fixed} ~~~~~({\rm Right-movers}) \\
\tfrac{\mless}{2} \, e^{-\theta} , ~~~\mless = m e^{-a}~ {\rm held ~ fixed} ~~~~~({\rm Left-movers}).
\end{cases}
\eeq
We will continue to label $\mless$ as $m$.  
This leads to Left and Right pseudo-energies $\vep_{L,R}$.   As explained in the last section,  and proposed in \cite{AL},  
the TBA equations for $\vep_{L}$ and $\vep_{R}$ are coupled with kernels that require factorizing $S_{\cdd}  = S_{LR} S_{RL}$,  which leads to
 \beq
\label{GLRpure} 
G_{RL} (\theta) = G_{LR} (-\theta) =  g  \, e^\theta /2 .
\eeq
The coupled TBA equations are as in \eqref{TBAmassless}.  

Due to the simple form of the kernels in \eqref{GLRpure} the pseudo-energies can now be expressed as 
\beq
\label{vepR}
\vep_R (\theta) = (1+ 2 h B_R ) \, r \, e^\theta /2, ~~~~~
\vep_L (\theta) = (1+ 2 h B_L ) \,  r \,e^{-\theta} /2,
\eeq
where $B_R$  only depends on $h$ and satisfies 
\beq
\label{BRinteq}
B_R   = \int_{-\infty}^ \infty \frac{d\theta}{2 \pi} \, \( r  e^\theta  /2 \) \, 
\log\[ 1 - e^{-(1+2 h\, B_R  )\, r e^\theta /2 } \] ,
\eeq
with  $B_L = B_R$ and $\vep_{L} (\theta)= \vep_R(-\theta)$.  
The central charge is 
\beq
\label{cmassless} 
c(h) = - \frac{6}{\pi} \( B_R (h) + B_L (h) \) = -\frac{12}{\pi} B_R(h) .
\eeq
The integral \eqref{BRinteq} can be done using
\beq
\label{convInt}
 \int_{-\infty}^\infty d \theta \, e^\theta \, \log \( 1- z  e^{-y \,e^\theta} \) = - \frac{ \Li_2 (z) }{y}, ~~~~ \Re (y) >0,
\eeq
where $\Li_2 (z) = \sum_{n=1}^\infty z^n / n^2$.   For complex $z$,  $\Li_2 (z)$ is the dilogarithm.  We only need $z= \pm 1$ 
 for bosons verses fermions and
$\Li_2 (1) = -2 \Li_2 (-1) = \zeta (1) = \pi^2/6$.
For $\sigma = +1$, 
we obtain the simple algebraic equation
\beq
\label{BalgEq}
B_R = - \frac{\pi}{12  (1+ 2 h B_R)} .
\eeq

In order to see that this is equivalent to the result  in \cite{AL},  which was obtained  in a  different manner directly from the massless TBA,  define
$B_R = - \frac{\pi}{12} \inv{A}$. 
Then 
$A$ satisfies the quadratic equation $ A = 1 - \pi h/(6A) $ as in \cite{AL}, which is easily solved.    
The result for $c$ is the simple expression \eqref{chGeneral} with $c_{IR} =1$.

\def\ghat{\hat{g}}

\def\psibar{\bar{\psi}}

Repeating the above analysis for fermions,  i.e. the non-critical Ising model,  one finds nearly identical formulas 
apart from a few signs.  One has 
\beq
\label{vepFermi}
\vep (\theta) = (1-h B )r \cosh \theta
\eeq
where now $B(r,h)$ satisfies the integral equation 
\beq
\label{BinteqFermi}
B = \int_{-\infty}^ \infty \frac{d\theta}{2 \pi} \, r \cosh \theta \, 
\log\[ 1 + e^{-(1-h\, B)\, r \cosh \theta } \] .
\eeq
The scale dependent $c$ is now given by 
\beq
\label{crhFermi}
c(r,h) =  \frac{6}{\pi} \, B(r,h) .
\eeq
The massless limit is also straightforward, namely, $B = B_R + B_L$ where $B_R (h) $ satisfies 
\beq
\label{BalgEqFermi}
B_R =  \frac{\pi}{24  (1- 2 h B_R)} .
\eeq
Solving this quadratic equation one finds
\beq
\label{Bsolved}
B_R (h) = \frac{\pi}{12}  \( 1 + \sqrt{1-\frac{\pi h}{3}} \)^{-1} , 
\eeq
and  $c(h)$ is as in \eqref{chGeneral} with $c_{IR} = 1/2$.

\section{Conclusions and Outlook}

Whereas pure $T\Tbar$ deformations are generally UV incomplete,  we have shown that such theories can be completed by including an infinite number of 
perturbations by more irrelevant operators.   Our study was based on the thermodynamic Bethe ansatz.      Consistent UV completions are expected to be both  rare  and not necessarily unique.    For instance, for the Ising model with $c_{IR}=1/2$, i.e. a free 
Majorana fermion, we found two such completions with $c_{UV} = 3/2$ and $7/10$,  both of which are $\CN=1$ supersymmetric.   The SUSY is broken in the IR and not at all 
anticipated;   rather this SUSY  becomes visible only after reconstruction and completion of the UV.     

Ultra-violet  incompleteness appears to be unavoidable for theories with $c_{IR} >0$ and  flows that are consistent with the c-theorem.   
Nevertheless, the pure $T\Tbar$ deformations can still be interesting,  possibly with applications as theories with a minimal, shortest  possible distance.   

We only considered in detail free conformal field theories in the IR since the main issues we attempted to understand are already present there.  
However the ideas presented in this paper can be extended to more complicated IR  CFT's.   This requires dealing with non-trivial Left-Left and Right-Right scattering
matrices $S_{LL}$ and $S_{RR}$ which are not simply equal to $\pm 1$,  and are appropriate to the massless IR CFT.   

It is commonly thought that irrelevant perturbations of an IR theory are intractable due to an infinite number of couplings and the irreversibility of RG flows from the UV to the IR.  
Perhaps the models considered in this paper provide some new lessons concerning the possibilities for  RG  flows 
since they provide counter examples to the commonly accepted properties of 
such flows.    For instance,  although in hindsight the flow of the tri-critical Ising model to the critical one was already known from the top down,  i.e. starting from the UV, 
we were able to ``rediscover"  this flow in the reversed direction,  i.e. starting purely in the IR.  We also found  a different completion with $c_{UV} = 3/2$ which was unanticipated.
  It is likely that symmetries play a large role in the possibility of such RG flows.

\section{Appendix:  Computing UV central charges from the TBA}

The calculation of  $c_{UV}$ from the TBA is a standard computation \cite{ZTBA,KM,Mussardo} which we briefly review. 

\subsection{Massive case}

Given a massive integrable model with a single bosonic or fermionic particle of mass $m$ with S-matrix $S(\theta)$,  kernel 
$G(\theta) = -i \d_\theta \log S(\theta)$,  and pseudo-energy $\vep (\theta)$, let us    define
\beq
\label{Ldef} 
L \equiv  -\sigma \log \(1-\sigma \, e^{-\vep} \).
\eeq
Then the TBA equation reads
\beq
\label{TBAmass}
\vep (\theta) = r  \, \cosh \theta  - (G \star L) (\theta) .
\eeq
Since $\vep$ and $L$ are even in $\theta$,  we can write 
\beq
\label{cTBA}
c (r) = \frac{6}{\pi^2} \int_{0}^\infty d \theta \, r \cosh \theta \, L(\theta)
\eeq
where $r=mR$.  
The UV central charge is $c_{UV} = \lim_{r \to 0} c(r)$.    The calculation of $c_{UV}$ is relevant to steps (i) and (ii) in Section IIIA.  

As $r\to 0$, $d\vep / d\theta \approx 0$ for $|\theta| \ll \log(2/r)$.   For this region of $\theta$, $L$ is approximately constant and $\vep$ is a constant $\vep_0$ which satisfies the transcendental equation 
\beq
\label{trans}
\vep_0 = \sigma k \log \( 1-\sigma e^{-\vep_0} \),
\eeq
where 
\beq
k \equiv \int_{-\infty}^\infty \frac{d \theta}{2 \pi} \, G(\theta).
\eeq
For most of the $\theta$ region of integration for positive $\theta$,  $m \cosh \theta \approx m\, e^\theta /2$,  thus the UV limit essentially corresponds to a massless theory 
with decoupled pseudo-energies $\vep_{L,R}$ which satisfy TBA equations with $S_{LL} = S_{RR} = S$.    One then has 
\beq
\label{cuvmassive}
c_{UV} = \lim_{r \to 0} \frac{6}{\pi^2} \, \int_0^\infty d \theta \, \, \tfrac{r}{2}  e^\theta  \, L(\theta)
\eeq
where 
$\vep$ now satisfies the massless TBA equation 
\beq
\label{veptilde}
\vep (\theta) = \tfrac{r}{2} e^\theta \, -  (G \star L) (\theta).
\eeq
Taking the derivative of the above equation and substituting, 
\beq
\label{cUV2}
c_{UV} = \lim_{r \to 0} \frac{6}{\pi^2} \, \int_0^\infty d \theta \[ \d_\theta \vep + \d_\theta (G \star L) \]  L(\theta).
\eeq
For the first term the integral over $\theta$ can be traded for an integral over $\vep$:
\beq
\label{cUV3}
c_1 = - \frac{6\sigma}{\pi^2} \int_{\vep_0}^\infty d \vep \, \log \( 1-\sigma e^{-\vep} \) =  \frac{6\sigma}{\pi^2} \,  \Li_2 (\sigma e^{-\vep_0} )
\eeq
where $\Li_2 (z) =  \sum_{n=1}^\infty z^n / n^2$ is the dilogarithm.   Integration by parts of the second term, assuming $G(\infty) = 0$,  gives a contribution that depends only on logs.  
The final result is 
\beq
\label{cUVfinal}
c_{UV} =  \frac{6\sigma}{\pi^2} \, \Lr_2 (\sigma e^{-\vep_0} ) 
\eeq
where $\Lr_2$ is the Rogers dilogarithm:
\beq
\label{rogers}
\Lr_2 (z) =  \Li_2 (z) + \tfrac{1}{2}  \log |z| \log (1-z) .
\eeq

\subsection{Massless case}

This case is similar to the massive case but not identical since the L,R pseudo-energies are now coupled.    
Let us write the TBA equations \eqref{TBAmassless} as follows
\beq
\label{TBAmasslessL} 
\vep_R = \tfrac{r}{2}  e^\theta  - G \star L_L , ~~~~~
\vep_L = \tfrac{r}{2} e^{-\theta}  - G \star L_R .
\eeq
where 
\beq
\label{Gs}
G_{RL} (\theta) = G_{LR} (\theta) = G (\theta) = G(-\theta)
\eeq
as in \eqref{Alltheta}. 
The symmetry of the kernel implies $\vep_L (\theta) = \vep_R (-\theta)$ and $c_L = c_R$. 
We can use this to express $c(r)$ as an integral over positive $\theta$ only:
\beq
\label{cless1}
c(r) = 2 \( c_R^{(+) }+ c_L^{(+)} \) 
\eeq
where $c^{(+)} = \int_0^\infty \cdots$.    As in the massive case, for $\theta \ll \log (2/r)$,  
$\vep_L = \vep_R = \vep_0$ where $\vep_0$ again satisfies the equation \eqref{trans}. 
For $\theta \gg \log (2/r)$,  $\vep_{R,L} \approx r e^{\pm \theta} /2$.   

Repeating the analysis of the massive case, one finds
\beq
\label{cless2}
2 c_R^{(+)} =   \frac{6\sigma}{\pi^2} \,  \Lr_2 \( \sigma e^{-\vep_0} \).
\eeq
For $c_L^{(+)}$ the limits of integration are different:
\beq
\label{cless3}
2 c_L^{(+)} = - \frac{6}{\pi^2} \int_0^{\vep_0} d \vep_L   \, L (\vep_L ) + \delta c_L^{(+)} 
\eeq
where $\delta c_L^{(+)}$ arises from the integration by parts which again produces the $\log \cdot \log $ term in $\Lr_2$. 
Replacing $\int_0^{\vep_0} = \int_0^\infty - \int_{\vep_0}^\infty$ one obtains
\beq
\label{cless4}
2 c_L^{(+)} =    \frac{6\sigma}{\pi^2} \( \Lr_2 \( \sigma e^{-\vep_0} \) - \Li_2 (\sigma) \).
\eeq
Putting this all together one gets
\beq
\label{cless5}
c_{UV} = \frac{6\sigma}{\pi^2} \( 2 \, \Lr_2 \( \sigma \, e^{-\vep_0} \) -  \Li_2 (\sigma) \). 
\eeq
Note that $6 \sigma \Li_2 (\sigma) /\pi^2 $ is  the central charge of the free theory which is subtracted in the final expression.


\begin{thebibliography}{99}



\bibitem{ctheorem}  A. B. Zamolodchikov, 
{\it Irreversibility of the flux of the renormalization group in a 2D field theory},
Pis'ma Eksp. Teor. Fiz. {\bf 43} (1986) 565. 


\bibitem{Weinberg} S. Weinberg,
{\it Critical Phenomena for Field Theorists},  In Zichichi, Antonino (ed.). Understanding the Fundamental Constituents of Matter. The Subnuclear Series. 14. pp. 1-52,
1978.  

\bibitem{ZTT}  A. B. Zamolodchikov, {\it Expectation value of composite field T anti-T in two-dimensional quantum field theory}, hep-th/0401146.

\bibitem{SZ}  F. A. Smirnov and A. B. Zamolodchikov, {\it On the  space of integrable quantum field theories}, Nucl. Phys. B 915 (2017) 363, arXiv:1608.05499.

\bibitem{Tateo1}
A Cavagli\`a, S Negro, I.M. Szecsenyi, R Tateo, 
{\it $T\Tbar$-deformed 2D quantum field theories}, JHEP 10 (2016) 112,
arXiv:1608.05534 [hep-th].



\bibitem{Tateo2}
R. Conti, L. Iannella, S. Negro, and R. Tateo, {\it Generalized Born-Infeld models, Lax operators and the $T\Tbar$ perturbation,} JHEP 11 (2018) 007, arXiv:1806.11515 [hep-th].

\bibitem{Rosenhaus}
Vladimir Rosenhaus, Michael Smolkin,
{\it Integrability and Renormalization under $T\Tbar$}, 
Phys. Rev. D 102, 065009 (2020), 
	arXiv:1909.02640 [hep-th]


\bibitem{Dubovsky} 
S. Dubovsky, V. Gorbenko, and M. Mirbabayi, {\it Asymptotic fragility, near AdS2 holography and $T\Tbar$, } JHEP 09 (2017) 136, arXiv:1706.06604 [hep-th].

\bibitem{Dubovsky2}
S. Dubovsky, V. Gorbenko, and G. Hernandez-Chifflet, {\it $T\Tbar$ partition function from topological gravity,} JHEP 09 (2018) 158, arXiv:1805.07386 [hep-th].

\bibitem{Cardy}
J. Cardy, {\it The $T\Tbar$ deformation of quantum field theory as random geometry,} JHEP 10 (2018) 186, arXiv:1801.06895 [hep-th].

\bibitem{Tateo3}
R. Conti, S. Negro, and R. Tateo, {\it The $T\Tbar$ perturbation and its geometric interpretation,} JHEP 02 (2019) 085, arXiv:1809.09593 [hep-th].

\bibitem{Verlinde}
L McGough, M Mezei, H Verlinde,  
{\it Moving the CFT into the bulk with $T\Tbar$},   
JHEP 10 (2018) 1, 
arXiv:1611.03470 [hep-th].


\bibitem{Hartman}
T. Hartman, J. Kruthoff, E. Shaghoulian, and A. Tajdini, {\it Holography at finite cutoff with a $T^2$ deformation,} ~JHEP 03 (2019) 004, arXiv:1807.11401 [hep-th].

\bibitem{Frolov}
S. Frolov, {\it $T\Tbar$ deformation and the light-cone gauge,}  arXiv:1905.07946 [hep-th].

\bibitem{Oku}  S. Okumura and K. Yoshida, 
{\it $T\Tbar$ perturbation and Liouville gravity},
Nucl. Phys. B957 (2020) 115083.

\bibitem{Jiang}  Y. Jiang, 
{\it Lectures on solvable irrelevant deformations of 2d quantum field theory}, 
arXiv:1904.13376 [hep-th]. 


\bibitem{AL}  A. LeClair,
{\it Thermodynamics of $T\Tbar$ perturbations of sinh-Gordon,  free boson,  and Liouville field theories},
arXiv:2105.08184 [hep-th]. 

\bibitem{Ebert} S. Ebert, H.-Y. Sun and Z. Sun, 
{\it $T\Tbar$ deformation in SCFTs and integrable supersymmetric theories}, 
arXiv:2011.07618 [hep-th].  

\bibitem{ZTBA}  
Al. B.  Zamolodchikov, {\it Thermodynamic Bethe Ansatz in relativistic models: scaling 3-state Potts and Lee-Yang models}, 
Nucl.  Phys.  B342 (1990) 695.

\bibitem{KM}  T. Klassen and E. Melzer,
{\it The thermodynamics of purely elastic scattering theories and conformal perturbation theory},
Nucl. Phys. B350 (1991) 635.

\bibitem{Mussardo}  G. Mussardo,  
{\it Statistical Field Theory.   An Introduction to Exactly Solved Models in Statistical Physics,}
Oxford University Press, 2010.   

\bibitem{Simon}  G. Mussardo and P. Simon, 
{\it Bosonic-type S-matrix, vacuum instability, and CDD ambiguity},
Nucl.Phys. B578 (2000) 527, 
arXiv:hep-th/9903072.


\bibitem{ZZ}   A. B. Zamolodchikov and    Al. B. Zamolodchikov,
{\it  Massless factorized scattering
and sigma models with topological terms,} 
Nucl. Phys. B379 (1992) 602.

\bibitem{Fendley}    P. Fendley, H. Saleur, Al.B. Zamolodchikov, {\it Massless flows II: the exact S-matrix approach}, Int. J. Mod. Phys. A 8
(1993) 5751,
arXiv:hep-th/930405.

\bibitem{Alyosha}  Al. B. Zamolodchikov,  
{\it From tri-critical Ising to critical Ising by thermodynamic Bethe Ansatz,}
Nucl.  Phys. {\bf B358} (1991) 524.


\bibitem{ZMinimalFlow}  A. B. Zamolodchikov,  
Sov. J. Nucl. Phys. 46 (1987) 1090.

\bibitem{CardyLudwig}  J. Cardy and A. Ludwig, 
Nucl.  Phys.  B285 (1987) 687.






\end{thebibliography}
\end{document}